\documentclass[twocolumn]{aastex63}
\pdfoutput=1

\usepackage{color}
\usepackage{mathtools}
\def\tess{\hbox{$TESS$}} 
\def\MEarth{\hbox{MEarth}}
\def\kepler{\hbox{$Kepler$}} 
\def\ktwo{\hbox{$K2$}} 

\def\numflares{\hbox{4}}

\usepackage{verbatim}

\newcommand{\msol}{M_\odot}

\newcommand{\E}[1]{\hspace{-0.01in}\times\hspace{-0.02in}10^{#1}}  % Scientific notation,
 
\newcommand{\angstrom}{\mbox{\normalfont\AA}}

\newcommand{\ha}{H\alpha}

\shorttitle{}
\shortauthors{Medina et al.}
\graphicspath{{./}{figures/}}
\begin{document}

\title{Flare Rates, Rotation Periods, and Spectroscopic Activity Indicators of a Volume-Complete Sample of Mid-to-Late M dwarfs within 15 Parsecs}

\correspondingauthor{Amber Medina}
\email{amber.medina@cfa.harvard.edu}

\author[0000-0001-8726-3134]{Amber A. Medina}
\affiliation{Harvard-Smithsonian Center for Astrophysics, 60 Garden St.,Cambridge, MA 01238}

\author[0000-0001-6031-9513]{Jennifer G. Winters}
\affiliation{Harvard-Smithsonian Center for Astrophysics, 60 Garden St.,Cambridge, MA 01238}

\author{Jonathan M. Irwin}
\affiliation{Harvard-Smithsonian Center for Astrophysics, 60 Garden St.,Cambridge, MA 01238}

\author[0000-0002-9003-484X]{David Charbonneau}
\affiliation{Harvard-Smithsonian Center for Astrophysics, 60 Garden St.,Cambridge, MA 01238}

\begin{abstract}

We present a study of flare rates, rotation periods, and spectroscopic activity indicators of 125 single stars within 15 parsecs and with masses between 0.1$-$0.3 $\msol$ observed during the first year of the \tess~Mission, with the goal of elucidating the relationship between these various magnetically connected phenomena. We gathered multi-epoch high resolution spectra of each target and we measured equivalent widths of the activity indicators Helium I D$_3$, $\ha$, and the Calcium infrared triplet line at 8542.09 $\angstrom$. We present 18 new rotation periods from \MEarth~photometry and 19 new rotation periods from \tess~photometry. We present a catalog of 1392 flares. After correcting for sensitivity, we find the slope of the flare frequency distribution for all stars to have a standard value of $\alpha$ = 1.98 $\pm$ 0.02.  We determine R$_{31.5}$, the rate of flares per day with energies above E = 3.16$\times$10$^{31}$ ergs in the \tess~bandpass. We find that below a critical value of $\ha$ EW =  -0.71 $\angstrom$, log R$_{31.5}$ increases linearly with increasing $\ha$ emission; above this value, log R$_{31.5}$ declines rapidly. The stars divide into two groups: 26\% have $\ha$ in emission, high flare rates with typical values of log R$_{31.5}$ = -1.30 $\pm$ 0.08, and have Rossby numbers $<$ 0.50. The remaining 74\% show little to no $\ha$ in emission and exhibit log R$_{31.5}$ $<$ -3.86, with the majority of these stars not showing a single flare during the \tess~observations.

\end{abstract}

\section{Introduction}

The magnetic field strength, rotation period, and age of main-sequence stars with outer convective envelopes are intimately linked.  Observations of magnetic phenomena such as coronal and chromospheric emission, star spots, and stellar flares have shown that as stellar rotation period increases, magnetic activity decreases \citep{Skumanich1972, Noyes1984,Kruse2010,Wright2011}. The increase in rotation period is believed to be the result of angular momentum losses via magnetized stellar winds. The stellar spin down leads to a decrease in dynamo activity and thus a decrease in overall magnetic activity as the star ages.  
 
For Sun-like stars with outer connective envelopes and radiative interiors, an $\alpha \Omega$ dynamo is believed to be responsible for the generation of a magnetic field \citep{Moffatt1978,Parker1979}. The key ingredient in this mechanism is the tachocline, the interface between the convective and radiative zones. In recent years, several studies have shown that fully convective stars, with masses less than 0.35 $\msol$, and lacking a radiative inner zone and hence a tachocline show a similar relationship of observable magnetic phenomena with rotation period \citep[e.g.][]{Newton2017,Wright2018}. These findings suggest either that the tachocline is not a central element in the generation of stellar magnetic field, or that fundamentally different field generation mechanisms are at work in these two classes of stars. Further observational studies of magnetic phenomena of fully convective stars aim to provide insights into dynamo processes of stars lacking a tachocline. 

Stellar flares are highly energetic transient events resulting from the re-connection of magnetic field lines in the stellar corona \citep{Martens1989}. These events emit radiation across the electromagnetic spectrum from radio to X-ray emission. White light flares, flares that exhibit broadband emission spanning from the near ultra-violet through to optical wavelengths, sometimes extending into the far ultra-violet and near infra-red have been studied extensively in recent years thanks to space-based dedicated photometric monitoring missions like \kepler/\ktwo. Several studies have focused on stellar flares and their relationship to properties of the stars, such as stellar rotation period, mass and age.  \citet{Davenport2016} used all short- and long-cadence data from the \kepler~mission to search for stellar flares. In their final sample of approximately 4000 flare stars, the majority of which are FGK and early M stars with masses above the fully convective limit, they find that the fraction of flaring stars increases with decreasing stellar mass. They also find that there exists a strong trend between flaring activity and rotation. They observe that stars with shorter rotation periods tend to flare more frequently and have more energetic flares than stars of the same mass with longer rotation periods. Coronal x-ray emission follows a similar trend showing an increase in the ratio of x-ray to bolometric luminosity, $L_x/L_{bol}$, as a function of decreasing Rossby number, defined as the rotation period divided by the convective turnover time, $\tau$, of the star. However in the case of x-ray emission,  $L_x/L_{bol}$ does not increase indefinitely at faster rotation periods. \citet{Vilhu1987,Pizzolato2003,Wright2011,Wright2016,Wright2018} show that $L_x/L_{bol}$ reaches a saturated value of 10$^{-3}$ at Rossby numbers below 0.13 for both partially and fully convective stars. Observations of saturated behavior are not confined to the stellar corona, they extend into the stellar chromosphere as well. \citet{France2018} demonstrates that far ultra-violet emission also shows saturation for FGK and early M stars with rotation periods less than 3.5 days. Emission of Calcium II H + K and $\ha$ also reach saturated values below a critical Rossby number \citep{Noyes1984,Douglas2014,Newton2017}. Although saturated behavior is observed in a number of magnetic activity indicators, it is unclear whether flares follow this trend \citep{Davenport2016,Howard2019}.

A study by \citet{Ilin2019} relates this decrease in the flaring activity observed by \citet{Davenport2016} in a more direct way to stellar age by using \ktwo~to examine flaring activity of late K - mid M dwarfs in clusters at different ages. They find a higher fraction of flare activity at younger cluster ages around 125 Myr than at older ages around 4 Gyr. This study is in line with previous age-activity-rotation relationship findings concerning stars above the fully convective boundary. 

In a series of papers \citet{Hawley2014,Davenport2014, Lurie2015,Silverberg2016} focus on fully convective stars observed by \kepler~short-cadence data to study flares on mid-to-late M dwarfs, most notably GJ~1234 and GJ~1245~AB. This paper series provides a wealth of findings concerning flare morphologies, and strong trends between flare energies, durations, amplitudes, and decay times. However, flaring behavior on a larger sample of fully convective stars is needed to probe relationships between flares and other properties of the star, such as rotation. Using existing ground based photometric observations from the \MEarth~survey, \citet{Mondrik2019} searched for flares in, at that time, the largest sample of photometric observations of mid-to-late M dwarfs. Their results indicate that rapidly rotating stars flare much more frequently than slowly spinning stars; their results further suggested that stars in the sparsely populated intermediate periods flared more frequently than either rapidly rotating stars or slowly spinning stars. However, \MEarth~is a ground based survey and thus cannot provide continuous observations, from which flare studies greatly benefit.

The Transiting Exoplanet Survey Satellite (\tess)~mission with its goal of finding planets around our nearest and/or brightest stellar neighbors is providing a large sample of mid-to-late M dwarfs for which a comprehensive study of flares is now possible.  \citet{Gunther2020} used the first two sectors of the primary mission to show that TESS provides an opportunity to study thousands of intrinsically faint mid-to-late M dwarfs at a photometric precision that was previously not possible at this scale. They show that 30\% of all mid-to-late M dwarfs are flaring compared to 1\% of FGK stars, further solidifying past findings; the fraction of magnetically active stars increases with decreasing stellar mass \citep{West2004,West2008,West2015}.

In this paper, we use the first 13 sectors of \tess~observations, which cover the first year of the mission and span most of the southern ecliptic sky to study flare rates and their relationship to other stellar properties such as rotation and chromospheric emission. We study a well-characterized sample of the nearest fully convective mid-to-late M dwarfs within 15 parsecs that have masses between 0.1 and 0.3 $\msol$. The stars comprise a robust statistical sample of nearby targets whose sizes and masses make them prime targets for near-future atmospheric studies of transiting terrestrial exoplanets; it is essential we understand the past and present stellar radiation environment. This sample also provides a means to probe further into the manifestations of magnetic activity on fully convective M dwarfs and how it compares to other observed relationships related to magnetic activity on both fully and partially convective stars. In \S~\ref{sec:sample} we describe the stellar sample. We describe the spectroscopic observations and analysis of our sample in \S~\ref{sec:spec_obs}. \S~\ref{sec:mearth_rot} outlines our measurement of new photometrically determined rotation periods using \MEarth~data. In \S~\ref{sec:tess}, we describe the photometric \tess~observations, rotation period measurements, and flare detection and analysis. We present the results in \S~\ref{sec:results}, and a discussion and conclusion follow in \S~\ref{sec:DC}.

\section{Stellar Sample}\label{sec:sample}
We begin with the volume complete sample of 419 main sequence stars with masses between 0.1$-$0.3$\msol$ and distances less than 15 pc from \citet{Winters2020b}. We remove all stars that were not observed by \tess~as part of its survey of the southern ecliptic sky in Year 1. All stars in the sample have either a trigonometric parallax from ground based monitoring or from the second data release of \textit{Gaia} \citep{GaiaDR22018}. \citet{Winters2020b} determined the masses of these stars using the K$_s$ magnitude and the relations presented in \citet{Benedict2016}. Using these relations, the typical uncertainties on the mass range from 4.7--14.0\%. These uncertainties are dominated by the scatter in the mass-luminosity relation. For this study we only use a subset of these stars, namely those that are single, as binaries that are unresolved in the \tess~photometry would prevent us from learning whether the flares, photometric rotation period, and/or chromospheric activity indicators all originate from the same source. We exclude all known double-lined and single-lined spectroscopic binaries, photometric eclipsing binaries, and visual binaries with separations of less than 63 arcseconds (3 \tess~pixels); the details of these multiple systems are noted in \citet{Winters2020b}. The exclusion of close binaries leaves 125 stars in our final sample.  We list the Names, TIC Identifiers, coordinates, masses, distances, \tess~magnitudes, as well as quantities derived in the following sections including spectroscopic activity indicators, rotation periods, Rossby numbers, and flare rates, in Table \ref{tab:MT}. We note that four stars in our sample are known to host transiting planets: GJ 1132 \citep{Berta-Thompson2015, Bonfils2018}, LHS 1140 \citep{Dittmann2017,Ment2019}, LHS 3844 \citep{Vanderspek2019}, and TOI-540 \citep{Ment2020}.

% put this monster table right here
\begin{deluxetable*}{lccl}
\tabletypesize{\scriptsize}
\tablecaption{Stellar Properties, Equivalent Widths, and Flare Rate (Table Format) \label{tab:MT}}
\tablehead{ 
\colhead{Column} & 
\colhead{Format} &  
\colhead{Units} & 
\colhead{Description}} 
\startdata 
1 & A22 & ... & Star Name \\
2 & A10 & ... & TIC Identifier \\
3 & F4.4 & hh:mm:ss.s & R.A in hours, minutes, seconds (J2000) \\
4 & F4.4 & dd:mm:ss.s & Declination in degrees, minutes, seconds (J2000) \\ 
5 & F1.2 & $\msol$ & Stellar mass \\
6 & F2.2 & parsecs & Distance \\
7 & F3.2 & days    & Photometric Rotation Period \\
8 & A1  & ... & Reference for Rotation Period \\
9 & F1.4 & mag & Semi-amplitude of variability \\
10 & F1.4 & mag & Uncertainty in semi-amplitude  \\
11 & F1.4 & ... & Rossby Number (Prot/$\tau$) \\
12 & F2.2 & mag & \tess~Magnitude \\
13 & F2.3 & $\angstrom$ & Equivalent Width of $\ha$ \\
14 & F2.3 & $\angstrom$ & Uncertainty in equivalent width of $\ha$ \\
15 & F2.3 & $\angstrom$ & Equivalent Width of Calcium II  \\
16 & F2.3 & $\angstrom$ & Uncertainty in equivalent width of Calcium II \\
17 &  F2.3 & $\angstrom$ & Equivalent Width of Helium I D$_{3}$  \\
18 & F2.3 & $\angstrom$ & Uncertainty in equivalent of  Helium I D$_{3}$ \\
19 & F2.2 & Log Flares day$^{-1}$ & Natural log rate of flares per day at energies greater than 3.16$\times 10^{31}$ ergs \\
20 & F2.2 & Log Flares day$^{-1}$ & Uncertainty in natural log rate of flares per day at energies greater than 3.16$\times 10^{31}$ ergs \\
\enddata 
\tablecaption{Full table available in machine-readable form.}
\end{deluxetable*}

\section{Spectroscopic Observations}
\label{sec:spec_obs}
As part of an ongoing effort initiated in September 2016 to gather multi-epoch high-resolution spectra of each star in the sample, we obtained two or more high-resolution spectra for each star. For stars with declinations below $-$15 degrees, we use the cross-dispersed, fiber-fed echelle CTIO HIgh ResolutiON (CHIRON) spectrogragh located on the CTIO/SMARTS 1.5 meter Telescope at CTIO. The CHIRON spectrograph covers the wavelength range 4100--8700 $\angstrom$ \citep{Tokovinin2013}. We used the image slicer mode for a resolving power of $R\approx80,000$. We acquired a thorium-argon hollow-cathode lamp spectrum through the science fiber both before and after every science spectrum. Exposure times ranged from 180s to 3~$\times$~1800s in good conditions, achieving a signal-to-noise ratio of 3-15 per pixel at $7150 {\angstrom}$. The spectra were reduced using the CHIRON pipeline described in \citet{Tokovinin2013}. 

We gathered spectra for all stars with declinations greater than 15 degrees with the high-throughput, fiber-fed, Tillinghast Reflector Echelle Spectrograph (TRES) located on the 1.5 meter Tillinghast Reflector at the Fred Lawrence Whipple Observatory on Mount Hopkins, Arizona. TRES has a resolution of R $\approx$ 44,000 and covers the wavelength range 3900-9100 $\angstrom$. Exposure times ranged from 120s to 3~$\times$~1200s to reach a signal to noise ratio of 3-25 at 7150 $\angstrom$. The spectra were reduced using the standard TRES pipeline \citep{Buchhave2010}.

\subsection{Spectroscopic Activity Indicators}
 Magnetic activity in mid-to-late M dwarfs can be monitored by tracking the strength of chromospheric features. We measured the equivalent widths (EWs) of the Helium I D$_3$ at 5875.6 $\angstrom$, $\ha$ at 6562.8 $\angstrom$, and one of the three Calcium infrared triplet lines at 8542.09 $\angstrom$. The other two Calcium lines at 8498.02 and 8662.14 $\angstrom$ are inaccessible in the CHIRON spectra due to breaks in the echelle orders. For consistency between the CHIRON and TRES spectra, we measure the same lines in each spectrum. 

In selecting the wavelength intervals that we use to evaluate the continuum, we take care to avoid regions contaminated by telluric features, as well as the molecular bands that are  in M-dwarf spectra. We measured EWs of the various features according to the standard equation,

\begin{equation}\label{eq:EW}
    \rm EW = \sum{\left( 1 - \frac{F(\lambda)}{F_c} \right)} \delta\lambda
\end{equation}

\noindent where F$(\lambda)$ is the flux in the EW region of interest and F$_c$ is the flux in the specified continuum region. In order to measure the flux contained within the limits of the feature window, we sum the flux in each pixel including fractional pixels. We measure the average flux in each continuum region on either side of the feature of interest and compute the average of the two continuum regions as F$_c$. The uncertainty in the equivalent width is the product of the fractional uncertainties in F$_c$ and F$(\lambda)$ added in quadrature and the measured EW value. We define our EW regions as well as continuum regions in Table \ref{tab:EW}. We take negative EW values to denote emission. As these activity indicators have been shown to vary as a function of time \citep{Lee2010,Kruse2010,Hilton2010}, especially during a large flare, we use the maximum EW value (least amount of emission) of our multi-epoch observations. All maximum equivalent width measurements are presented in Table \ref{tab:MT}.

\begin{center}
\begin{deluxetable}{lccc}
\tablecaption{Regions used for Measurement of Equivalent Widths \label{tab:EW}}
\tablehead{ 
\colhead{Feature} & 
\colhead{F$_{C1}$} & 
\colhead{F($\lambda$)} &   
\colhead{F$_{C2}$}  \\
& 
\colhead{$\angstrom$} & 
\colhead{$\angstrom$} &
\colhead{$\angstrom$}}
\startdata
$\ha$         &  6554.1 $-$ 6559.1 & 6560.3 $-$ 6865.3 & 6566.5 $-$ 6570.5  \\
He I D$_3$  & 5870.0 $-$ 5873.0 & 5874.6 $-$ 5876.6   & 5877.6 $-$ 5880.6 \\
Calcium II    & 8537.0 $-$ 8540.0 &   8541.3 $-$ 8542.8  & 8560.0 $-$ 8580.0 \\
\enddata

\end{deluxetable}
\end{center}

\section{New Photometric Rotations Periods From MEarth}\label{sec:mearth_rot}

MEarth-North and \MEarth-South arrays each comprise eight 40 cm telescopes and are located on Mt. Hopkins in Arizona and at Cerro Tololo Inter-American Observatory (CTIO), Chile, respectively \citep{Nutzman2008,Irwin2015}. 
 
 Stellar rotation periods can be deduced from photometric modulations resulting from stellar spots rotating into and out of view. Rotation periods for 35 stars in our sample were previously published \citep[e.g, 33 from][]{Newton2016,Newton2018}, one from \citet{Suarez2016}, and one from \citet{Vanderspek2019}; these ranged from 0.1-140 days. We present additional rotation periods from \tess~photometry in Section \ref{sec:tess}, however, in general, rotation periods longer than 10 days are inaccessible to \tess~due to the 14 day spacecraft orbit; longer periods are accessible to \MEarth. We follow the same methodology used in \citet{Newton2016,Newton2018} to estimate the rotation periods of 1 additional star with \MEarth-North and 17 additional stars with \MEarth-South; these were largely stars where observations were not yet complete at the time \citet{Newton2016,Newton2018} were published. We list the new rotation periods (ranging from 65-180 days) in Table \ref{tab:MT}.

\section{TESS Photometry}\label{sec:tess}
\begin{figure*}
\includegraphics[scale=.70,angle=0]{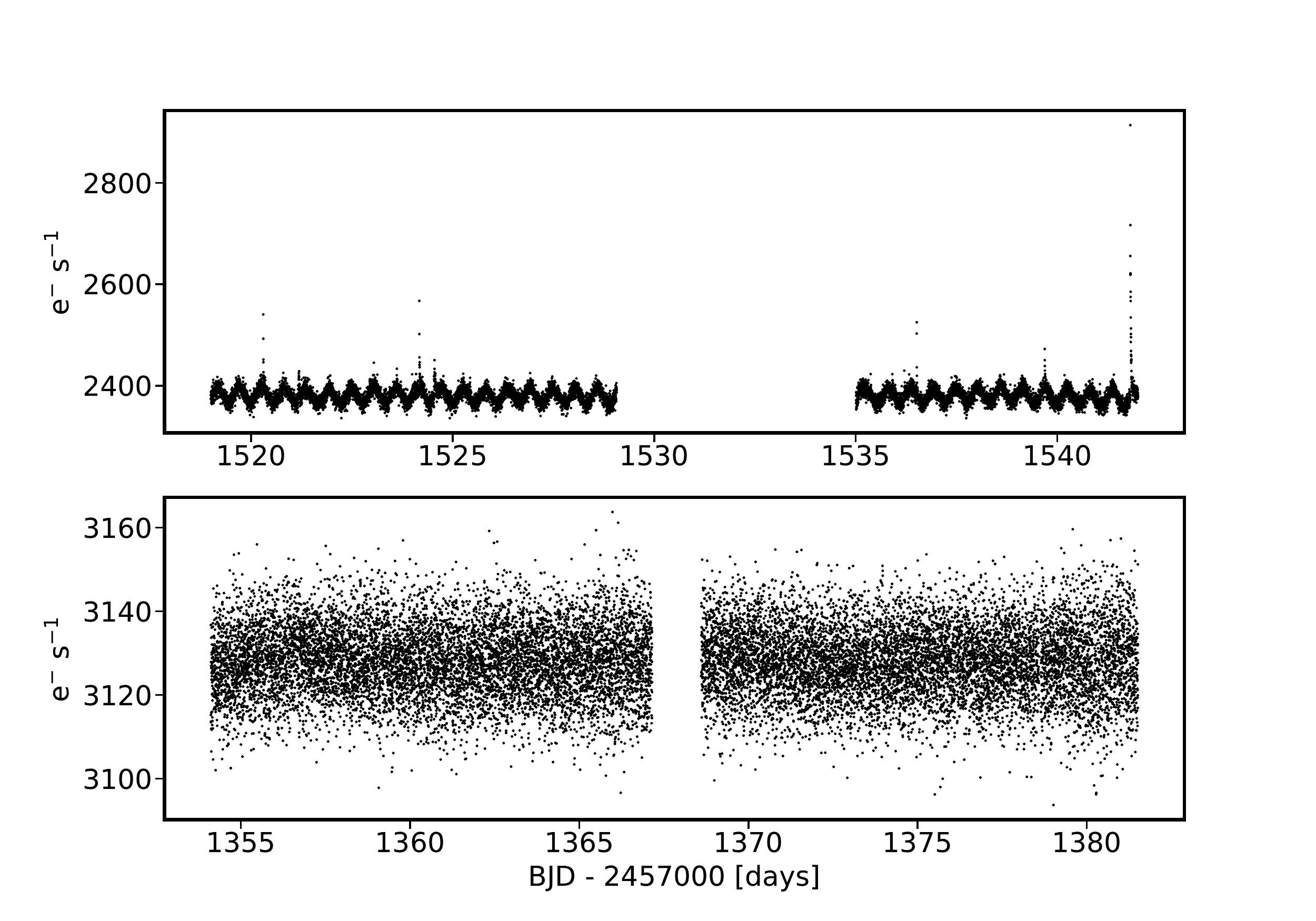}
%\vspace{-1cm}
%\hspace{-0.53cm}
%\includegraphics[scale=1.5,angle=0]{panoplyline_3.png}
\caption{\tess~PDCSAP light curves for 2 mid-to-late M dwarfs showing the variety of photometric variability we see in our stellar sample. The top panel is SCR J0914-4134 with a rotation period of 0.55 days. The bottom panel shows SCR J0246-7024 which has a rotation period of 106.80 days. \label{fig:stars_example}}
\end{figure*}

As part of its survey of the southern ecliptic hemisphere in Year 1, \tess~obtained two-minute cadence data of the 125 stars in our sample through a guest investigator program (PI Winters; G011231).  In Figure \ref{fig:stars_example}, we show examples of \tess~light curves for two stars in our sample, SCR J0914-4134 and SCR J0246-7024 with rotation periods of 0.55 and 106.80 days, respectively. These two stars are each representative of the two broad categories of photometric behavior we observe: stars showing an abundance of photometric variability and stars with little to no photometric variability.

\subsection{Photometric Rotation Periods with TESS}
\begin{figure}[ht]
\includegraphics[scale=.5,angle=0]{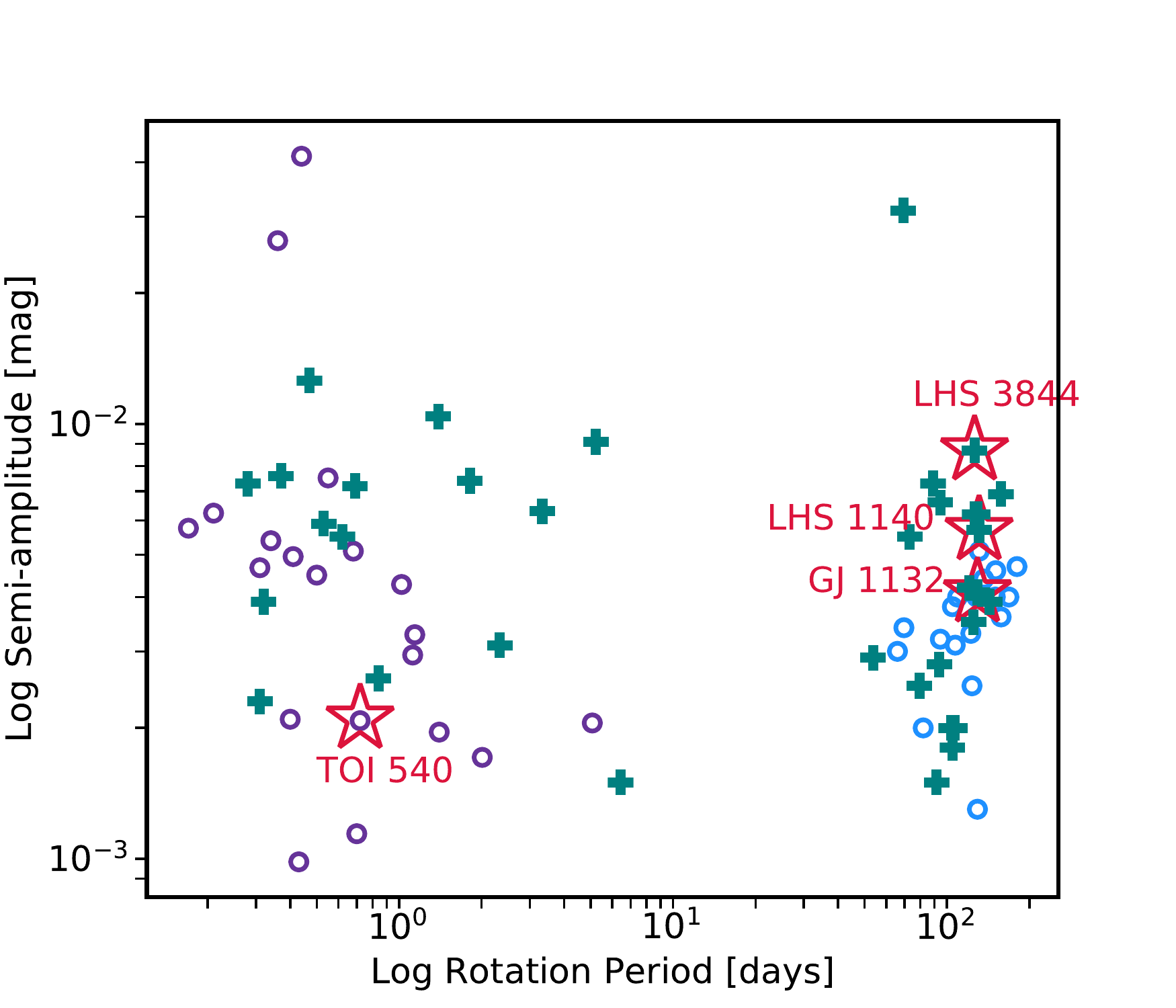}
%\vspace{-1cm}
%\hspace{-0.53cm}
%\includegraphics[scale=1.5,angle=0]{panoplyline_3.png}
\caption{The semi-amplitude of variability as a function of rotation period. The teal plus symbols shows previously published rotation periods; all but two of these come \citet{Newton2016,Newton2018}. The blue circles show new rotation periods measured in this work using data from \MEarth. The purple circles show the rotation periods measured in this work using the \tess~data. The red stars outline known transiting planet host stars. \label{fig:prot_hist}}
\end{figure}

We measured 19 new rotation periods ranging from 0.17--5.07 days using the \tess~photometry. To measure rotation periods, we use a periodogram analysis as outlined in \citet{Irwin2011,Newton2016, Newton2018}. We search for periods from 0.05 - 14 days. Given that \tess~observes a given sector for 27 days, we exclude periods that could not be observed for two full cycles. We do not attempt to measure rotation periods longer than 14 days for stars with multiple sectors of data. All of the rotation periods we measured with \tess~are less than 6.0 days, which is likely due to the observing baseline of one orbit of the \tess~spacecraft being 13.7 days followed by a multi-hour interruption as the data are down-linked. In Figure \ref{fig:prot_hist} we show the semi-amplitude of variability as a function of the rotation period for all stars with measured photometric rotation periods measured either in this work using \tess~or \MEarth~ as well as those previously published. The distribution of periods shows two populations, with no stars having intermediate periods between 6.43 and 53.74 days. We find no significant difference in the amplitude of the variation between the short period and long period populations, nor between the amplitudes found by \MEarth~and those found by \tess. We find consistency between previously published rotation periods for stars with periods less than or equal to 6.43 days presented in \citet{Newton2016, Newton2018}. We list rotation periods, semi-amplitudes of variability, and the uncertainties on the semi-amplitudes in Table \ref{tab:MT}.

\subsection{Stellar Flares with TESS}\label{sec:flare_alg}
We use the \tess~two-minute cadence data to search for flares, determine their energies, and ultimately tabulate the rate and energy distribution of flares for each star.

\subsubsection{Flare Detection Algorithm}
\begin{figure*}
\includegraphics[scale=.70,angle=0]{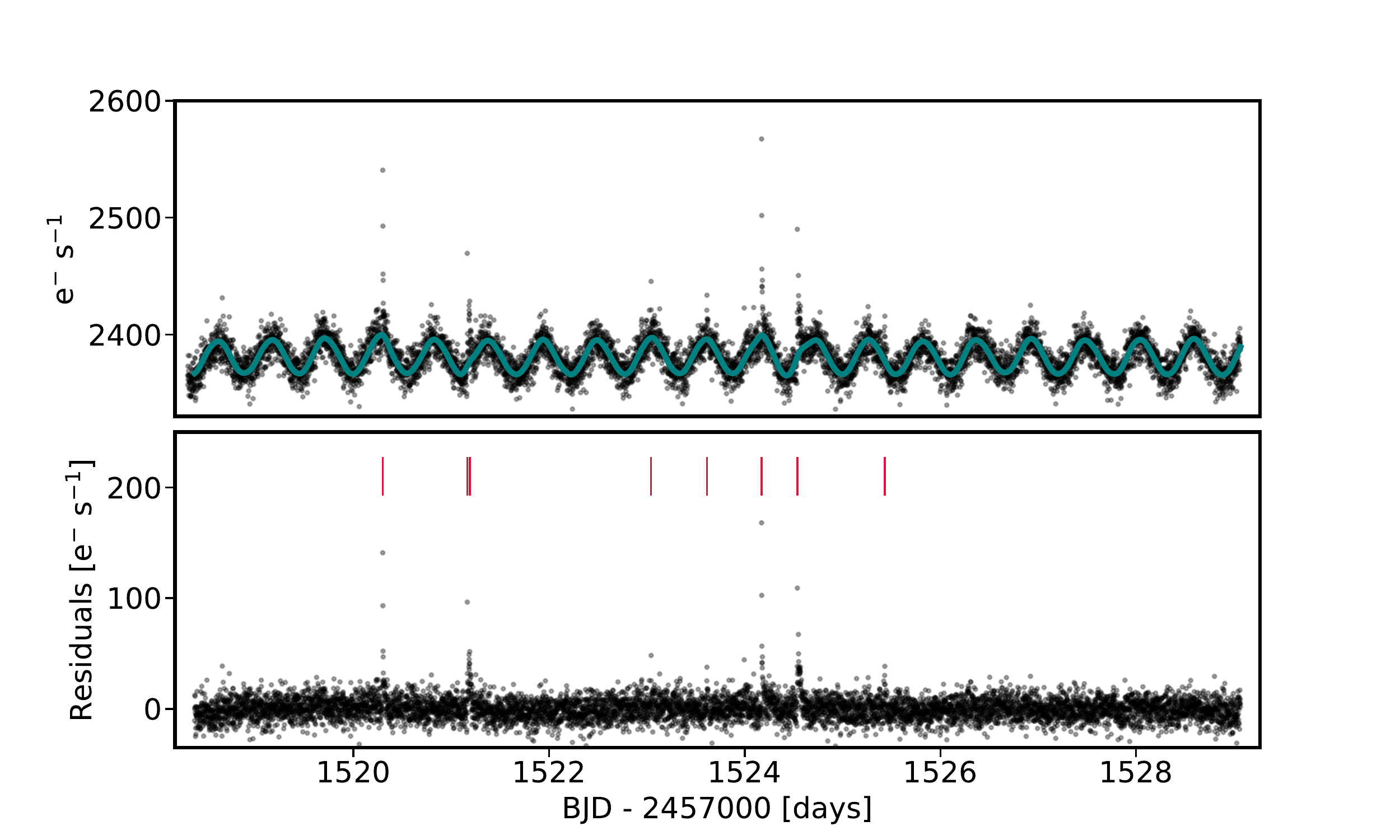}
%\vspace{-1cm}
%\hspace{-0.53cm}
%\includegraphics[scale=1.5,angle=0]{panoplyline_3.png}
\caption{Example of our detrending and flare finding algorithm with SCR J0914-4134. The PDCSAP light curve of this star is also shown in Figure \ref{fig:stars_example}. The top panel shows the \tess~light curve with the GP model we use to detrend the data (teal curve) for one orbit of a \tess~sector. The bottom panel shows the residuals with flares found by our flare finding algorithm shown with a red tic mark.\label{fig:detrend}}
\end{figure*}

In order to detect flares in the \tess~light curves, each light curve must be detrended to remove  variability such as rotational modulation due to stellar spots.  We use the Pre-Search Data Conditioning Simple Aperture Photometry (PDCSAP) light curve. We use all data with the quality flags equal to zero or 512. Quality flag 512 denotes a point that is an impulsive outlier. Upon examination of the light curves, we noticed points labeled with quality flag of 512 where being removed largely from within flares. We decided to include these points as removing them would lead to an underestimation of flare energies. We define light curve segments as windows of continuous observations with no interruptions exceeding two hours. We model each light curve segment individually. We used the python package {\sc exoplanet}  \citep{exoplanet:exoplanet,Foreman2017} to model the light curve with a Gaussian Process (GP) taking the form of a combination of two simple harmonic oscillators. The GP kernel is the sum of two simple harmonic oscillators shown in Equations \ref{eq:sho1} and \ref{eq:sho2}:

\begin{equation}
\label{eq:sho1}
    SHO_1(\omega_{GP}) = \sqrt{\frac{2}{\pi}} \frac{S_1\,\omega_1^4}
        {(\omega_{GP}^2-{\omega_1}^2)^2 + {\omega_1}^2\,\omega_{GP}^2/Q_1^2}
\end{equation}

and

\begin{equation}
\label{eq:sho2}
       SHO_2 (\omega_{GP}) = \sqrt{\frac{2}{\pi}} \frac{S_2\,\omega_2^4}
        {(\omega_{GP}^2-{\omega_2}^2)^2 + {\omega_2}^2\,\omega_{GP}^2/Q_2^2}
\end{equation}

where, 

\begin{equation}
    Q_1 = 0.5 + \tau_3 + \tau_4
\end{equation}

\begin{equation}
    \omega_1 = \frac{4\pi Q_1}{\tau_2 \sqrt{4Q_1^2 - 1}}
\end{equation}

\begin{equation}
    S_1 = \frac{\tau_1}{\omega_1 Q_1}
\end{equation}

\begin{equation}
    Q_2 = 0.5 + \tau_3
\end{equation}

\begin{equation}
    \omega_2 = \frac{8\pi Q_2}{\tau_2 \sqrt{4Q_2^2 - 1}}
\end{equation}

\begin{equation}
    S_2 = \frac{\tau_5~\tau_1}{\omega_2 Q_2}
\end{equation}

\noindent where $\tau_1$ is the  amplitude of variability, $\tau_2$ is the primary period of the variability, $\tau_3$ describes the quality factor of the oscillator, $\tau_4$ describes the difference between the quality factors of the first and second modes of the two oscillators,  $\tau_5$ describes, the fractional amplitude of the secondary mode to the primary mode, and $\tau_6$ describes a jitter term added to account for excess noise. We use maximum likelihood optimization to determine these parameters for each light curve. We subtract the GP model from the PDCSAP light curve. We search these residuals for flares according to the criteria outlined in \citet{Chang2015} and used in other stellar flare studies by \citet{Davenport2016,Ilin2019}. We define a flare as 3 consecutive 3$\sigma$ positive deviants. An example of our flare finding algorithm and detrending procedure can be seen in Figure \ref{fig:detrend}.  

\begin{figure*}
\includegraphics[scale=.70,angle=0]{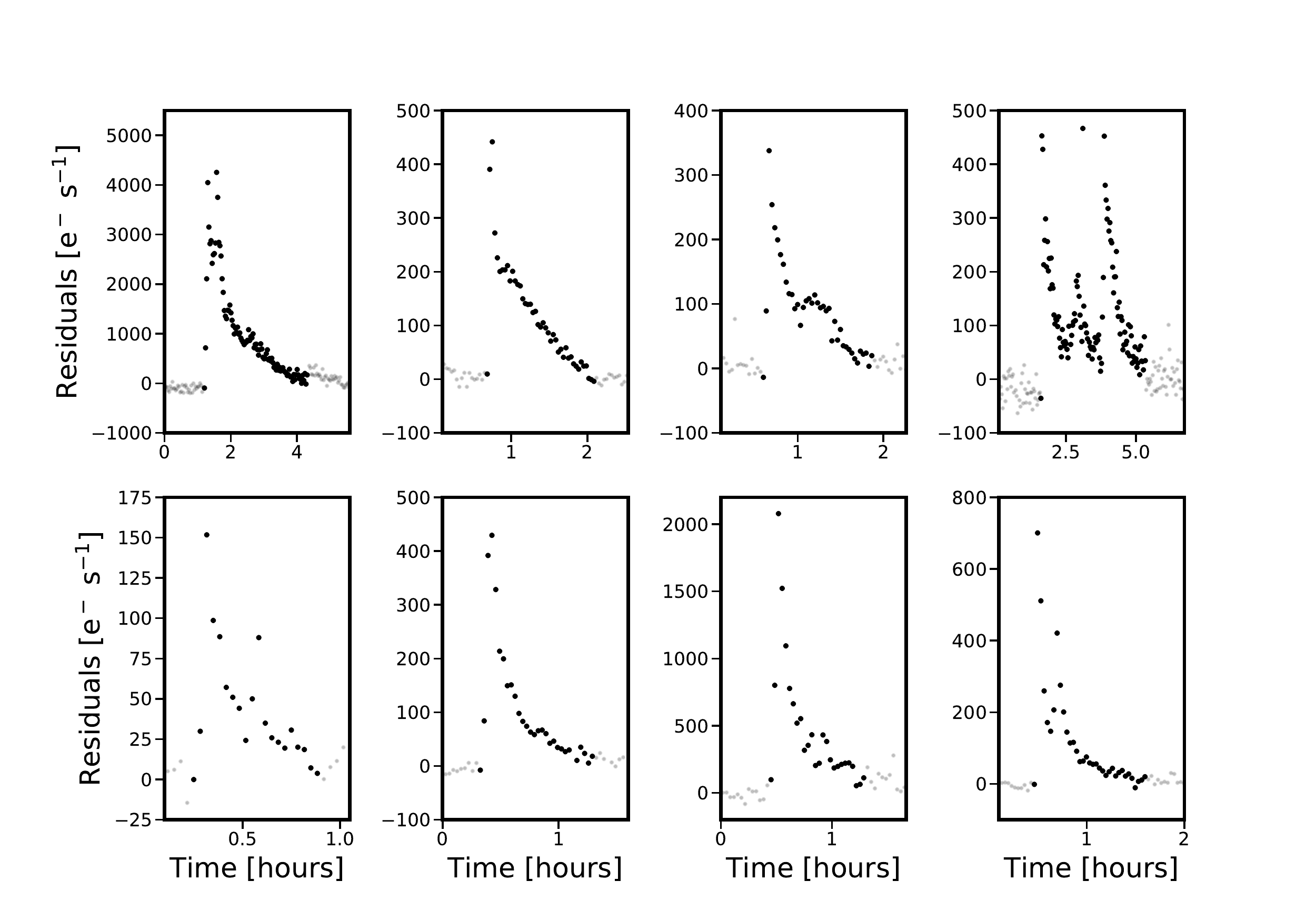}
%\vspace{-1cm}
%\hspace{-0.53cm}
%\includegraphics[scale=1.5,angle=0]{panoplyline_3.png}
\caption{Examples of flares detected with our flare finding algorithm for 6 different stars in our sample. Flares come from Proxima Centauri, SCR~J0754-3809, UPM~J0409-4435, APMPM~J0237-5928, L~87-10, and  Gl~54.1. The black points denote points used to calculate the flare energy.   \label{fig:flare_gallery}}
\end{figure*}

We find at least one flare for 75 out of 125 stars.  In total we find 1392 flare events. We find classical single peaked flares as well as multi-component complex flares, but for the purpose of this analysis we treat multi-component flares as a single event as we are only concerned with measuring their integrated flux. In Figure \ref{fig:flare_gallery}, we show several examples representing the range of flare phenomena observed.

\subsubsection{Flare Energies}
%We determine flare energies by integrating the area under the curve of each detected flare, the distance to the star, the effective area of the \tess ~telescope, and the average energy of a photon in the \tess ~bandpass given by \ref{eq:FE}. We decided to determine the energy of the flare in the \tess ~bandpass instead of the bolometric energy of the flare. Other studies present flares on an absolute scale by assuming the flare is a black body with a temperature of 10,000 \citep{Gunther2020, Shibayama2013}. Although we realize the value of having our flares presented on a absolute scale we note a few challenges with as summing a constant black body temperature for the flare. \citet{Davenport2019} points out that flares have been shown to display excess blue and red emission, and whether or not the flare shows a constant temperature in its flaring state is not well characterized. 

%\begin{equation}
%\label{eq:FE}
%    E  = \frac{\left(\sum_{i = t_{start}}^{t_{stop}}{F_i~dt} \right)4 \pi D^2 <E_{\gamma}>}{A_{\rm eff}}
%\end{equation}
%Where $F_i$ is the flux at time $t_i$ and summed from the time start of the flare, $t_{start}$ to the time at which the flare stops, $t_{stop}$. D is the distance to the star in centimeters, $<E_{\gamma}>$ is the average energy of a photon in the \tess bandpass evaluated at 785.5 nanometers, and A$_{\rm eff}$ is the effective area of the \tess~telescope valued at 86.6 cm$^2$ \cite{Ricker2015}. 

The wavelength dependence of the emission from stellar flares is not completely observationally constrained \citep{Davenport2012,Kowalski2013,Kowalski2016,Loyd2018,MacGregor2020}, nor is it constant from one flare to the next. Ignoring line emission and modeling the flare as a Planck function at T=9000K, we find by integrating over the Planck curve from 600--1000 nm, that the \tess~bandpass captures 20.9\% of the flare energy. Hence our rate R$_{31.5}$, which corresponds to the number of flares per day above 3.16 $\times$ 10$^{31}$ ergs would correspond to the rate of flares per day with a bolometric energy of 1.51$\times$10$^{32}$ erg. For comparison, we note that Carrington-like events which are the largest flares occurring on the Sun in modern times have energies ranging from a few $\times$ 10$^{32}$ -- 10$^{33}$ ergs \citep{Carrington1859,Hodgson1859,Woods2004,Emslie2012,Moore2014}. We emphasize that we have no constraints on the spectral energy distribution of the flares we have studied with \tess; modeling these as simple blackbodies with a uniform temperature is surely incorrect \citep[e.g.][]{Kowalski2010,Kowalski2013,Kowalski2016,Kowalski2019} and hence we choose to present our entire analysis with the energies quantified in the \tess~bandpass.

We determine flare energies using the quiescent luminosity of the star in the \tess~bandpass, $L_{T}$, the distance to the star, and equivalent duration, ED. Distances were determined using a trigonometric parallax from the second data release of \textit{Gaia} \citep{GaiaDR22018} with the exception of one star, LHS~3746 for which no parallax was listed in DR2. For this star, we used the parallax measured from ground-based monitoring \citep{Henry2006}. The equivalent duration is defined as the time over which the quiescent star radiates the same amount of energy as was emitted by the flare \citep{Gershberg1972,Huntwalker2012, Hawley2014,Davenport2016}. Functionally, the ED is measured by integrating the area under the flare in the light curve of the residuals, e.g. after subtracting the photometric baseline modeled with the Gaussian Process. The flare energy is calculated by multiplying the equivalent duration, measured in seconds, by the quiescent luminosity of the star.  We determine $L_T$ using the zero-point magnitude flux of \tess~(4.03 $\times$ 10$^{-6}$~ergs~cm$^2$~s$^{-1}$; \citep{Sullivan2015}), the \tess~magnitude, $T_{mag}$, and the distance to each star. We calculate $T_{mag}$ from $I_{KC}$ and $K_s$ colors using a relation developed by Guillermo Torres and presented in \citet{Winters2019}. The largest source of uncertainty in our flare energies is the determination of where the flare ends. We quantified the uncertainty by decreasing and increasing the duration of the detected flare by subtracting and adding an additional point to the end of the flares, calculating the flare energies, and examining the ratio of those energies to the measured flare energies presented in Table \ref{tab:flares}. We found that the mean change in the flare energies was 5\%, and we assigned this uncertainty to all our flare energies. For one star, Gl~54.1, 28 minutes of data were missing from the PDCSAP light curve between Barycentric Julian dates (BJD) of 2458390.2510450 - 2458390.2704895 that overlapped with what we believe to be a large flare. Upon examination of the Simple Aperture Photometry (SAP) light curve, we found the missing PDCSAP data were preceded by a large flux increase followed by an exponential decay reminiscent of the classic flare morphology. We investigated the missing timestamps in the PDCSAP and SAP light curves for five other stars in Sector 3 observed with the same camera. We found that for all five stars, the same 28 minutes of data were missing from each PDCSAP light curve. We examined the target pixel file and found that the potential flare appeared to coincide, simultaneously, with a cosmic ray, thus corrupting the data. We posit that the bright flare and potential cosmic ray observed in Gl~54.1 with peak time recorded at BJD of 2458390.2579894 in the SAP light curve disrupted the calibration of the other stars around it and thus those points were marked with poor quality and removed from the PDCSAP light curves altogether.  The removal of this data means that for this particular flare with a peak time of 2458390.2704895 listed in Table \ref{tab:flares}, we underestimate the true flare energy, but we do not attempt to increase the flare energy to account for the missing data. We did not observe any other instances of data being removed from a flare.

\subsubsection{Flare Completeness Correction}\label{sec:flare_C}
Our sensitivity to flares in the \tess~data is dependant upon the intrinsic luminosity of the star \citep{Lacy1976}, the intrinsic energy of the flare, the distance to the star, the degree to which the light curve was affected by instrumental systematics, the variability of the star itself, and the ability of our algorithm to account for that variability. For each star we compute a completeness function. The completeness function is computed by injecting flares at \tess~band energies ranging from $1\times 10^{28}$ - $1\times10^{33}$ ergs. We chose 30 logarithmically (log$_{10}$) spaced energies from this range. We use the flare template described in \citet{Davenport2014} to inject flares into each light curve at each energy. We note that the \citet{Davenport2014} template does not account for all possible types of flare morphologies on stars such as sympathetic flaring \citep{Hawley2014,Davenport2014} or flares that show a gradual increase in the impulsive phase of the flare like the one seen in \citet{Kowalski2016}. However, we are only concerned with determining the energy of the injected flare, which does not depend on the morphology of the flare. As long as these flares show 3 positive 3$\sigma$ outliers, our flare finding algorithm would be sensitive to them. Each flare is parameterized by the time of peak flux, the full width at half maximum (FWHM), and the amplitude of the flare. We injected ten flares at a given energy into each light curve with care not to overlap with real flare events. We then deployed our flare finding algorithm and recorded the events that were recovered. We repeated this ten times for a total of 100 flares injected into each light curve at each energy. A flare was deemed recovered if the peak time was within 14 minutes of the injected flare peak time, and if the amplitude was within 20\% that of the amplitude of the injected flare.  

\begin{figure*}
\includegraphics[scale=.7,angle=0]{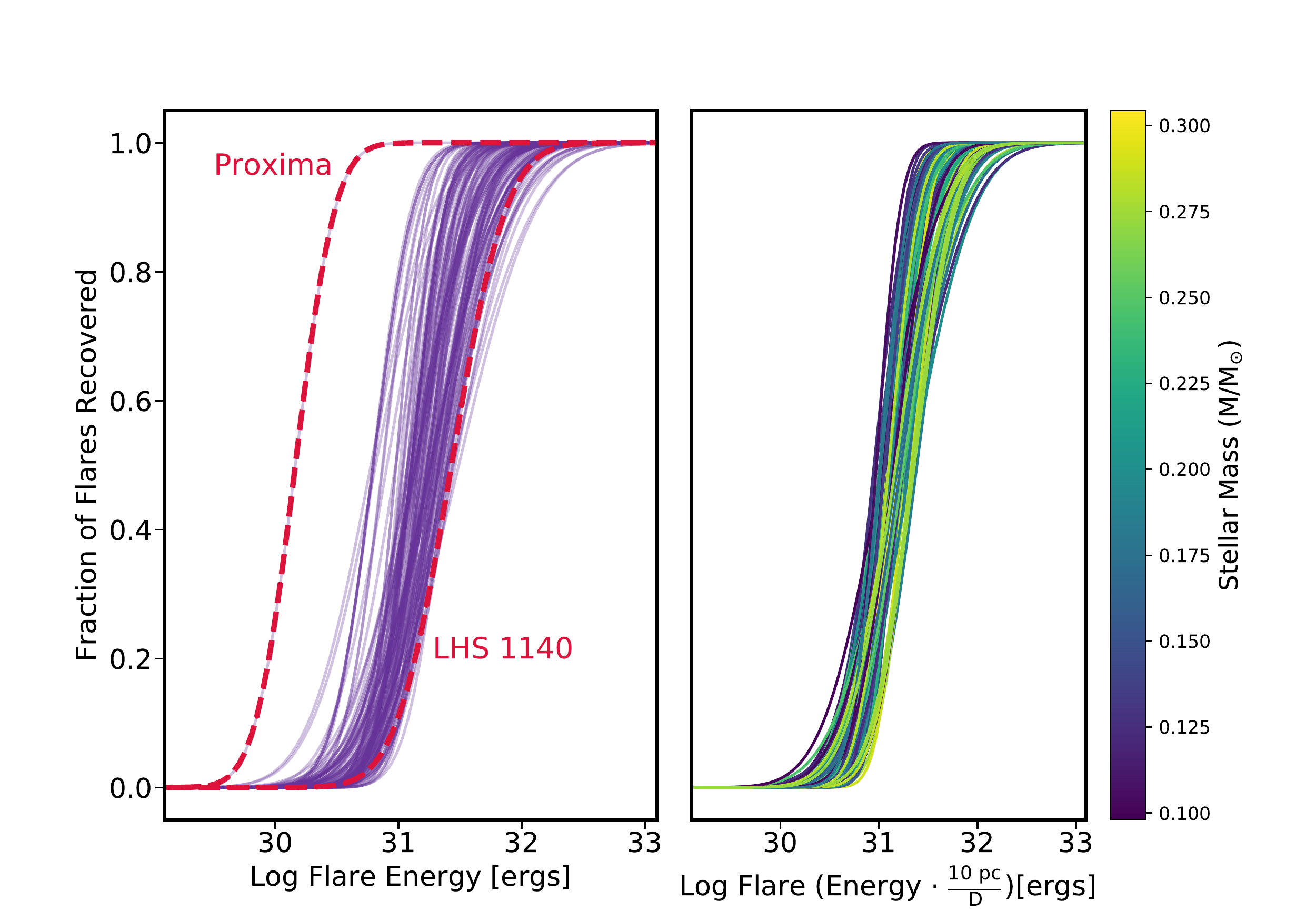}
%\vspace{-1cm}
%\hspace{-0.53cm}
%\includegraphics[scale=1.5,angle=0]{panoplyline_3.png}
\caption{Left panel shows, for each target star, the completeness function, which is the fraction of flares recovered by our flare finding algorithm as a function of energy (purple solid lines). Dashed red lines show the completeness function for the closest star in our sample, Proxima Centauri at 1.30 parsecs (left dashed line) and for the farthest star in our sample, LHS~1140, at 14.99 parsecs (right dashed line). The right panel shows the completeness function for each star in our sample scaled to a distance of 10 parsecs. The stellar mass is represented by the line color indicated by the colorbar.}  \label{fig:comp_func}
\end{figure*}

We model the completeness function, which is the fraction of injected flares recovered as a function of the $\rm{\log}$ of the flare energy, as an error function,

\begin{equation}\label{eq:erf}
    P(x) = \frac{1 + erf[k(x - b)]}{2}
\end{equation}

\noindent where x is equal to the natural log of the flare energy, and k and b are constants. A note that we use log in this paper to denote natural log; log base 10 will be denoted at log$_{10}$. We show the modeled completeness function for all stars in Figure \ref{fig:comp_func}, as well as the modeled completeness functions scaled to a distance of 10 parsecs. We show in Figure \ref{fig:comp_func} that the mass of the star can play a role in the flare energies recovered as lower energy flares are more easily detected on lower mass stars due to a greater contrast between the flare and the intrinsic luminosity of the star. In addition, \citet{Lacy1976} shows that it is also the case that stars with higher masses also intrinsically produce higher energy flares. However, we find the energy at which flares are recovered in our study predominantly depends on the distance to the star. The mass range in our study is smaller than the range of stellar masses presented in \citet{Lacy1976}.
An example of this can be seen in Figure \ref{fig:comp_func} where the energy at which flares begin to be detected in Proxima Centauri, which is at a distance of 1.30 parsecs, is an order of magnitude lower in flare energy than for the most distant star in our sample, LHS~1140, at 14.99 parsecs. 

We present the time of peak flux, amplitude, and duration for each flare in Table 3. We compare the flares that we find with the flares found in \citet{Gunther2020}, who report flares for the first two sectors of \tess~observations. We have 10 stars that overlap with their sample. We recover all 124 flares for these 10 stars and we detect an additional 90 flares that are not listed in Table 1 of \citet{Gunther2020}.

%\begin{deluxetable*}{lccccccc}
%\tabletypesize{\scriptsize}
%\tablecaption{Catalog of Stellar Flares \label{tab:flares}}
%\tablehead{ 
%\colhead{Star Name} & 
%\colhead{TIC ID} &   
%\colhead{\tess~Magnitude} &   
%\colhead{Time of Peak Flux} & 
%\colhead{Flare Amplitude} &
%\colhead{Duration} &
%\colhead{Equivalent Duration} &
%\colhead{Flare Energy} & \\
%\colhead{} & 
%\colhead{} & 
%\colhead{(mags)} & 
%\colhead{(BJD)} &
%\colhead{(counts/second)} & 
%\colhead{(seconds)}  &
%\colhead{(seconds)}  &
%\colhead{(ergs)} &
%&&
%} 
%\startdata 
%L87-10 & 234526939 & 11.16 & 2458355.0874 & 51.35 & 1440.0 & 6.23 & 1.61e+31 \\
%L87-10 & 234526939 & 11.16 & 2458357.0402 & 55.35 & 840.0 & 4.38 & 1.13e+31 \\
%L87-10 & 234526939 & 11.16 & 2458359.5471 & 117.27 & 2520.0 & 22.80 & 5.87e+31 \\
%\enddata 
%\tablenotetext{*}{A full version of this table is available in machine readable format}
%\end{deluxetable*}

% put this monster table right here
\begin{deluxetable*}{lccl}
\tabletypesize{\scriptsize}
\tablecaption{Catalog of Stellar Flares \label{tab:flares} (Table Format)}
\tablehead{ 
\colhead{Column} & 
\colhead{Format} &  
\colhead{Units} & 
\colhead{Description}}
\startdata 
1 & A22 & ... & Star Name \\
2 & A10 & ... & TIC Identifier \\
3 & F2.2 & mag & \tess~Magnitude \\
4 & F7.4 & BJD & Time of Peak Flux \\ 
5 & F3.2 & cts s$^{-1}$ & Flare Amplitude \\
6 & F4.1 & seconds & Flare Duration \\
7 & F2.2 & seconds  & Equivalent Duration \\
8 & E1.2  & ergs & Flare Energy in the \tess~Bandpass \\
\enddata 
\tablecaption{Full table available in machine-readable form.}
\end{deluxetable*}

\subsubsection{Flare Frequency Distribution}
The flare frequency distribution (FFD) describes the rate of flares as function of energy and follows the probability distribution shown in Equation \ref{eq:FFD} \citep{Lacy1976},

\begin{equation}\label{eq:FFD}
    N(E)dE = \Omega~E^{-\alpha}dE
\end{equation}

\noindent where $\alpha$ is the slope of the power law and $\Omega$ is a normalization constant. \citet{Ilin2019} in a recent study using \ktwo~light curves of K and M stars, posit that $\alpha$ may be the same for all stars; a value of approximately 2. A value of $\alpha$ $\geq$ 2 has important implications for coronal heating as it provides an answer to whether the number of flares, especially those at lower energies (10$^{24}$ --10$^{25}$ ergs) are abundant enough to heat the quiescent corona \citep{Audard2000,Gudel2002}. If $\alpha$ $<$ 2, the energy supplied to the corona depends on the most energetic events; low energy flares do not occur frequently enough to deposit the energies necessary to account for coronal heating \citep{Hudson1991}. However as the power law steepens to a value of 2 or greater, lower energy flares are plentiful enough that they deposit the energy necessary to heat the corona. \citet{Doyle1985} found that the time averaged U-band energy of flares is proportional to and is about an order of magnitude smaller than the quiescent coronal x-ray emission for a sample of early-to-mid M dwarfs. Similarly, they find that the total radiation emitted by a flare is also about an order of magnitude greater than the U-band energy of the flare. Flares may account for coronal heating if the bolometric radiation in very small flares has the same energy partition in x-ray and UV as larger flares \citep{Doyle1985,Osten2015}. However whether flares can fully account for the observed coronal temperatures has not yet been determined; measuring $\alpha$ can help determine this.

In past studies, the value of $\alpha$ was typically determined with a linear fit to the cumulative number of flares per day as a function of energy in log-log space where the exponent becomes $\alpha$ - 1. However \citet{Clauset2009} makes the case that this formalism is not appropriate for a number of reasons. The errors on $\alpha$ and $\Omega$ are difficult to estimate in the case of a linear regression because the the noise is not Gaussian in the log of the cumulative number of flares at each value of the log of the flare energy. \citet{Clauset2009} also argue that fits to linear regressions typically do not satisfy conditions concerning normalization of the probability distribution and thus cannot be correct. We chose to follow the formalization outlined in \cite{Clauset2009} for our FFD analysis with a modification that accounts for the flare completeness correction described in \S~\ref{sec:flare_C}. We begin by fitting $\alpha$ for only those stars that show 5 or more flares above the energy at which the completeness function was at least 30\%; 38 stars meet this criterion. 

The likelihood we use to fit for the slope of the FFD, $\alpha$, must take into account the fact that our flares come from two different distributions. One is the observed underlying power-law that stellar flares obey, and the other is the completeness function. This $\rm log$ likelihood takes the form:

\begin{multline*}
    \rm Log~\mathcal{L} = \rm N_{\rm fl} \rm \log~\Omega + \sum_{i=1}^{N_{\rm fl}} \left(1-\alpha \right)x_i \\ + \sum_{i=1}^{\rm N_{fl}} \rm log \left[\frac{1 + erf\left(k\left(x_i-b\right)\right)}{2} \right]\\
    + \rm N_{\rm fl} \rm \log \lambda - \lambda - \rm log (N_{fl} !) 
\end{multline*}

\noindent where $\alpha$ is the slope of the FFD, x$_{i}$ is the natural log of the energy of flare $i$, $\Omega$ is a normalization constant required so that the integral of the probability density function evaluates to one, b and k are parameters describing the error function used in the completeness correction (Equation \ref{eq:erf}), and as we assume flare events follow a standard Poisson Process, the third term includes the number of observed flares N$_{\rm fl}$ and $\lambda$, the rate of flares.

\begin{figure}[ht]
\includegraphics[scale=.55,angle=0]{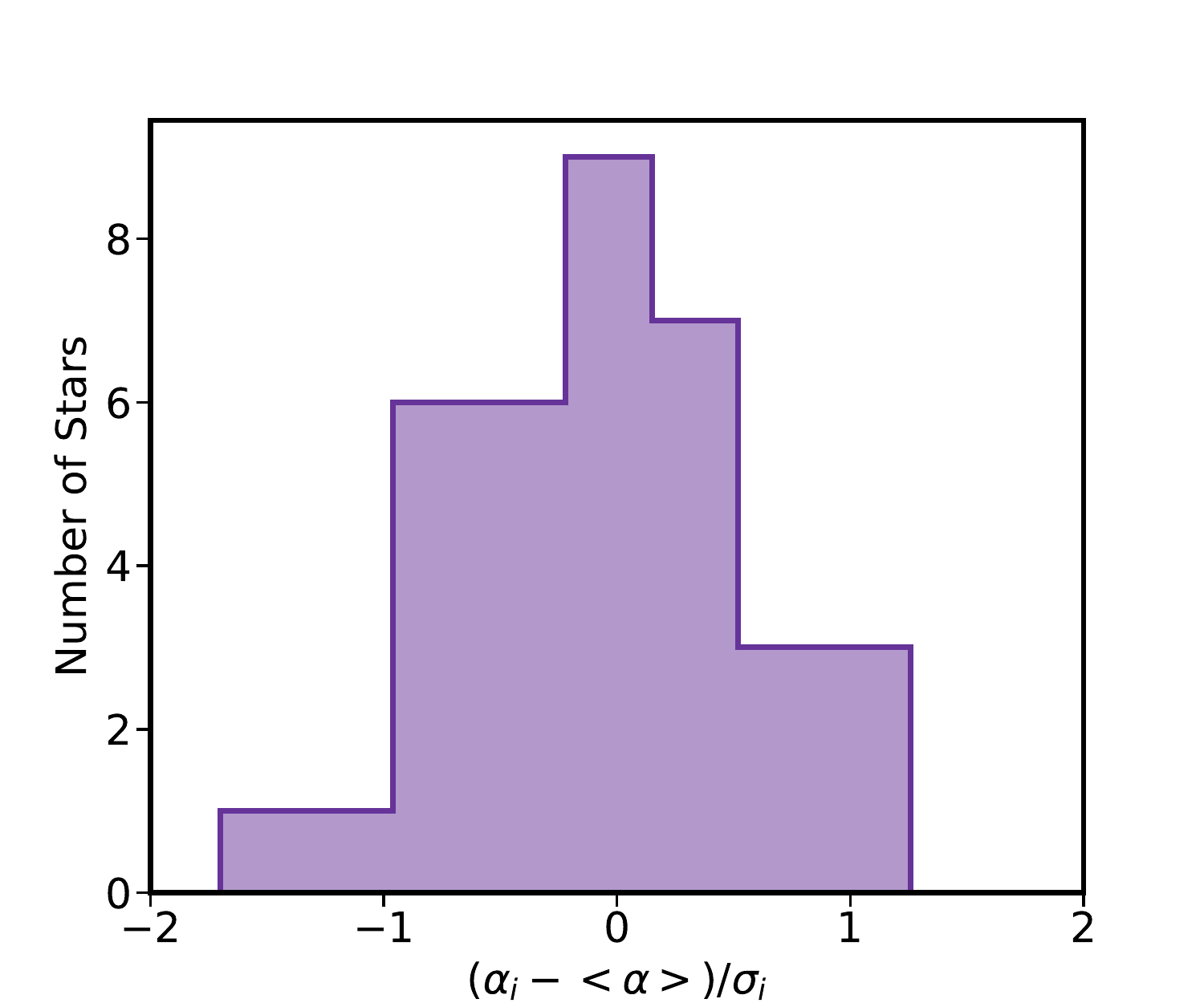}
%\vspace{-1cm}
%\hspace{-0.53cm}
%\includegraphics[scale=1.5,angle=0]{panoplyline_3.png}
\caption{Distribution of the Z-score ($\alpha_i - <\alpha>$)/$\sigma_i$ for fitted values of $\alpha$, the slope of the flare frequency distribution, where $\alpha_i$ are the best fit values of the slope and $\sigma_i$ are  the uncertainties on the slope estimates. \label{fig:alpha_hist}}
\end{figure}

We use Markov Chain Monte-Carlo methods with the python package \textit{emcee} \citep{Foreman2013} for the determination of $\alpha$ and the expected rate of flares, $\lambda$. We chose to use a Jeffreys prior on $\alpha$ and a uniform prior on $\lambda$. The distribution of $\alpha$ values for the 38 stars that have 5 or more flares is shown Figure \ref{fig:alpha_hist}. 

The results in Figure \ref{fig:alpha_hist} show that all 38 stars are consistent with the same value for the FFD slope $\alpha$. We compute a weighted average of the $\alpha$ values to obtain a value of 1.98 $\pm$ 0.02; consistent with the value of 2, which is the minimum sufficient to account for coronal heating. Because high energy events are intrinsically rare, and because of the relatively short baseline observations of \tess~ the inference of the rate carries large Poisson counting errors relative to long baselines like those of \kepler. We have accounted for this effect with the log likelihood we have employed in this analysis. Further, \citet{Davenport2020} compares \tess~observations spanning 2 sectors and \kepler~observations which span 11 months for mid M GJ~1243. They find that although \tess~observed fewer rare high energy events than \kepler~ (owing to its much shorter baseline), the $\alpha$ value measured using the \tess~data is consistent with the value measured using the \kepler~data.

Having established a common value of $\alpha$, we can fix this value to estimate the flare rate for stars with fewer than 5 flares, and to obtain a more precise value of the flare rate for stars with 5 or more flares. Using a fixed $\alpha =$ 1.98, we determine the flare rate (or, an upper limit, for stars with no flares) for all 125 stars. We then computed the natural log rate of flares above a certain flare energy denoted by log R, following the definition of \citet{Davenport2019}.  We chose the threshold flare energy to be 3.16$\times 10^{31}$ ergs (log$_{10}$(E) = 31.5) in the \tess~bandpass as this is the energy at which the completeness function is equal to 50\% or above for all stars in our sample. The R$_{31.5}$ flare rate and the calculated uncertainty on the flare rate are listed in Table \ref{tab:MT}. We note that stars that have many flares either because they are intrinsically very active or stars that have many sectors of observations typically have a smaller uncertainty in their measured flare rates.

\section{Results}\label{sec:results}

\begin{figure*}
\includegraphics[scale=.6,angle=0]{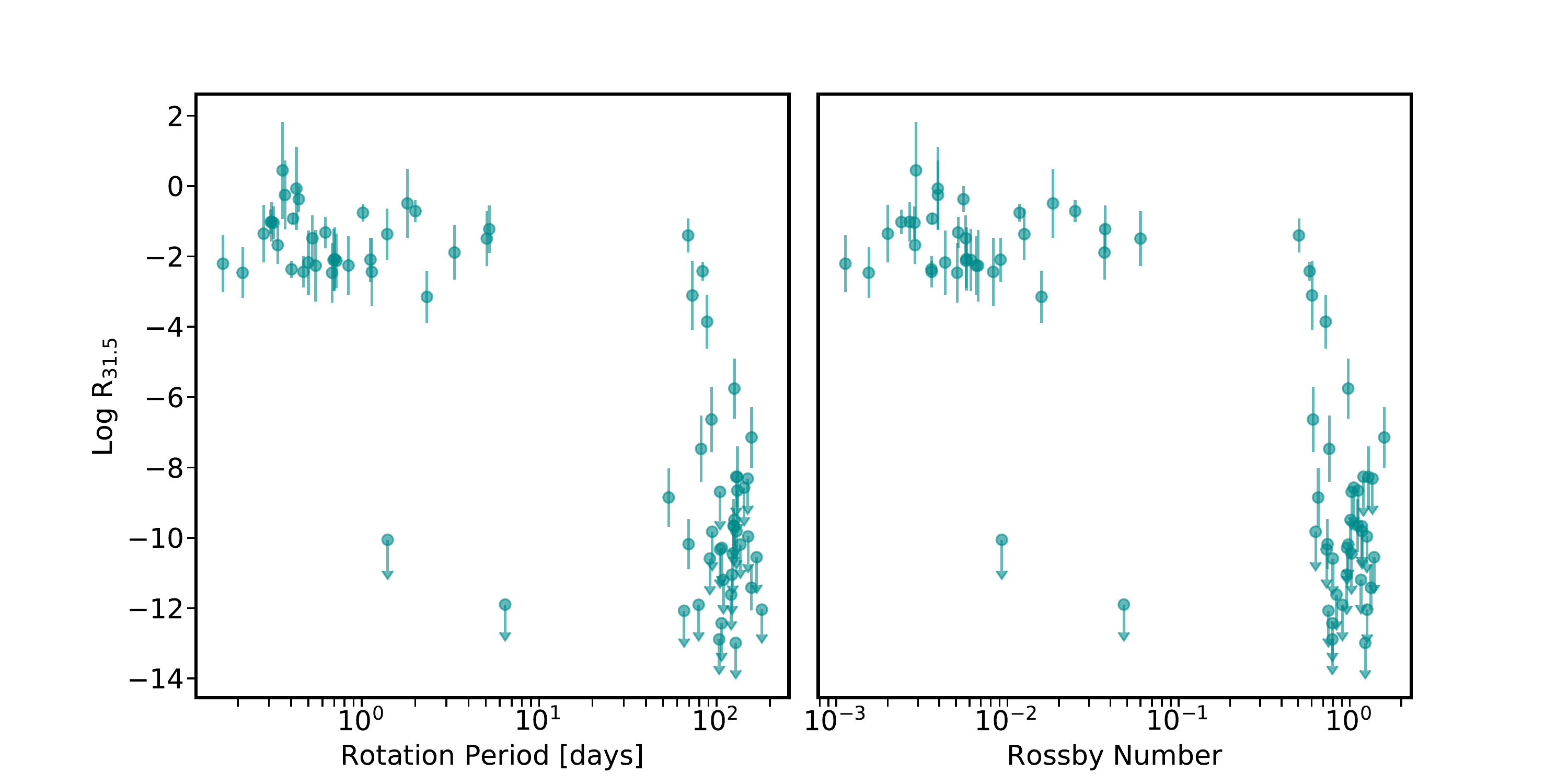}
%\vspace{-1cm}
%\hspace{-0.53cm}
%\includegraphics[scale=1.5,angle=0]{panoplyline_3.png}
\caption{The left panel shows the log number of flares per day above our completeness threshold of E = 3.16$\times$10$^{31}$ ergs as a function of the stellar rotation period. The right panel shows the same flare rate as a function of Rossby number. \label{fig:prot_fl}}
\end{figure*}

%The empirical relationship presented in \citet{Wright2018} was established by determining the period at which the coronal x-ray emission becomes saturated in a given V-K color range (used as a proxy for mass).

\subsection{Flare Rate and Stellar Rotation Period}
 Figure \ref{fig:prot_fl} shows the rate of flares, R$_{31.5}$, as a function of stellar rotation. A common way to examine the effects of rotation and its relationship to other stellar properties while marginalizing over stellar mass is through the use of Rossby number, Ro, which is defined as the rotation period of star divided by its convective turnover time. We calculate convective turnover times using the stellar mass and the empirical relationship from \citet{Wright2018}. We list the Rossby numbers for stars with a measured rotation period in Table \ref{tab:MT}. In the right panel of Figure \ref{fig:prot_fl}, we show the flare rate as a function of Rossby number. Because the stars in our sample span a fairly narrow mass range, the plots in the right and left panels of Figure \ref{fig:prot_fl} are very similar. 
 
Figure \ref{fig:prot_fl} shows a clear saturated relationship between the flare rate and Rossby number. In other studies of magnetic activity where saturation occurs, the saturation is only observed if the activity metric is normalized by the bolometric luminosity of the star. \citet{Lacy1976} shows that the flare rate at a given energy is dependent upon stellar luminosity in that low-mass stars overall produce lower energy flares than higher mass stars. After applying our sensitivity completeness correction, we find no dependence of the flare rate upon the stellar mass or luminosity, for either the saturated or unsaturated regimes. However, our study probes a narrower range of stellar masses than that of \citet{Lacy1976}. We determine the weighted average of the saturated flare rate for all stars with Ro $<$ 0.1. The flare rate shows a constant value of log -1.30 $\pm$ 0.08 flares per day below Ro = 0.1, and rapidly declines with increasing Rossby number above Ro = 0.5. Two notable outliers in Figure \ref{fig:prot_fl} are SCR~J1855-6914 and 2MASS~J23303802-8455189 with rotation periods of 1.40 and 6.43 days, respectively and log R$_{31.5}$ values less than -9.00 flares per day; we exclude these stars from the calculation of the saturated value. We posit that these stars may have unknown binary companions, leading to incorrect mass estimates erroneously placing them in our sample. The presence of a binary companion could also mean that the rotation period and flare limits correspond to different members of the binary.

\subsection{Flare Rate and Spectroscopic Activity Indicators}
The observation and characterization of flares is a powerful tool for developing an intuition for the evolution of the magnetic field and is essential to understanding the habitability of their planets. However, as flares are stochastic phenomena \citep{Lacy1976}, the probability of observing a flare on a faint mid-to-late M dwarf during an a typical nightly observation is low, especially from the ground, so determining a flare rate for these stars is difficult. Thus, it is would be useful to determine a relationship, should one exist, between the flare rate and other activity indicators that are readily accessible with a spectroscopic observation. 

\begin{figure*}
\includegraphics[scale=.65,angle=0]{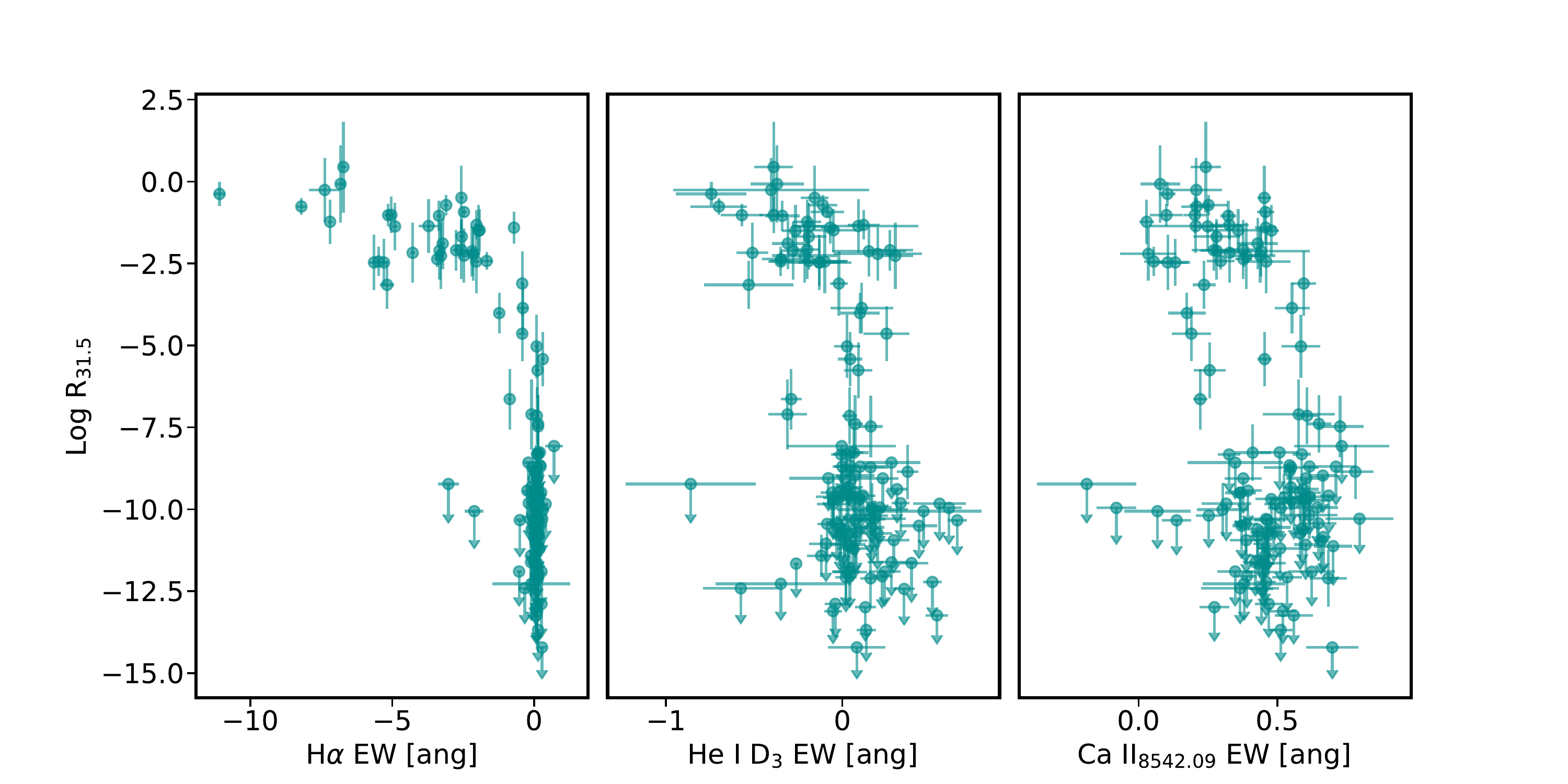}
%\vspace{-1cm}
%\hspace{-0.53cm}
%\includegraphics[scale=1.5,angle=0]{panoplyline_3.png}
\caption{The log rate of flares per day with energies above E = 3.16$\times$ 10$^{31}$ ergs as a function of measured equivalent widths of $\ha$ (left), He I D$_3$ (middle), and the calcium infrared triplet line at 8542.1 $\angstrom$ (right). Upper limits are shown as downward pointing arrows. \label{fig:EWs}}
\end{figure*}

In Figure \ref{fig:EWs} we show the relationship between equivalent widths of $\ha$, Ca II, and He I D$_3$ lines and R$_{31.5}$. We computed the Pearson correlation coefficients and p-values, which indicate the probability of the correlation being produced by an entirely uncorrelated data set for $\ha$, Ca II, and He I D$_3$ features to be -0.58 (p-value = 6.65e-05), -0.32 (p-value = 0.04), and -0.32 (p-value = 0.05) respectively. The equivalent width of $\ha$ has the strongest correlation with the flare rate. We observe a clear distinction between the population of stars with $\ha$ in emission and saturated flare rates, and stars with no $\ha$ emission and low flare rates. In contrast, for He I D$_3$ and Ca II, the average values between the saturated flare rate, and low flare rate populations overlap substantially. Thus, measuring these quantities will not allow one to deduce which regime a given star is in. We will focus on $\ha$ for rest of the analysis. 

In the left panel of Figure \ref{fig:EWs},  we observe a clear relationship between flare rate and $\ha$ that is similar to the relationship that exists for various chromospheric and coronal activity indicators as a function of Rossby number observed in \citet{Noyes1984,Wright2011,Douglas2014,Newton2017,Wright2018}. Stars that have an $\ha$ equivalent width value less than -0.7 angstroms have a higher flare rate relative to those that have measured EW values greater than -0.7 angstroms.

We look into the relationship between the EW of $\ha$, measured in angstroms, and the log of the flare rate by fitting a piece-wise function to all stars that show \numflares~or more flares. The functional form is shown in Equation \ref{eq:PW_3} and will be referred to as H$_1$, where x$_{\ha}$ is the $\ha$ EW, $\gamma$ is the slope for EW values below the critical value, $\beta$ is the slope above the critical EW value, and C is the rate at which the two regimes intersect.  We determine the critical $\ha$ value, X$_{\ha_c}$ between where the flare rate begins to increase linearly towards higher flare rates using $\chi^2$ minimization fitting Equation \ref{eq:PW_3} at each EW value and taking the X$_{\ha_c}$ value giving the smallest $\chi^2$. X$_{\ha_c}$ has a value of -0.71 angstroms.

\begin{equation}\label{eq:PW_3}
\rm log~R_{31.5}(x_{\ha}) = 
        \begin{cases}
            \gamma (x_{\ha} - X_{\ha_c}) + C &x_{\ha} \leq X_{\ha_{c}}\\
            \beta  (x_{\ha} - X_{\ha_c}) + C &x_{\ha} > X_{\ha_{c}}\\
        \end{cases}
\end{equation}

\begin{figure*}
\centering
\includegraphics[scale=.63,angle=0]{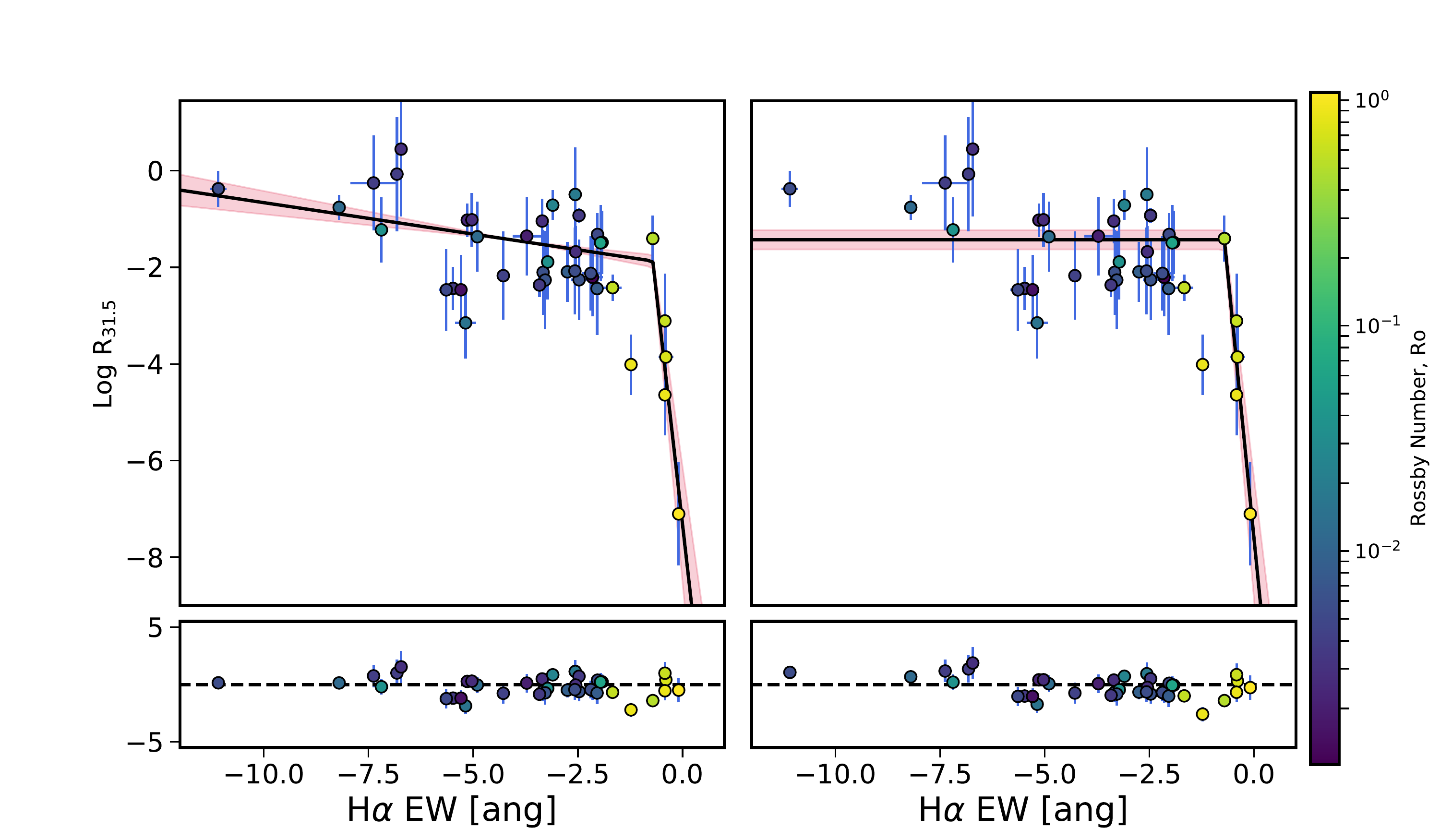}
%\vspace{-1cm}
%\hspace{-0.53cm}
%\includegraphics[scale=1.5,angle=0]{panoplyline_3.png}
\caption{Both panels show the log rate of flares per day with energies above E = 3.16$\times$ 10$^{31}$ ergs as a function of measured equivalent widths of $\ha$ for all stars in our sample with \numflares~or more flares and a measured rotation period. The color of the points indicates the value of the Rossby number shown by the colorbar. The black line shows the best fit model assuming a model that assumes two separate linear regimes(left panel), Equation \ref{eq:PW_3}) and a saturated relation (Equation \ref{eq:PW} (right panel). The red shaded region shows the 1$\sigma$ confidence interval. \label{fig:ha_fit}}
\end{figure*}

Using the critical value, we use Markov-Chain Monte-Carlo methods and the python package \textit{emcee} \citep{Foreman2013} to determine the parameters C, $\gamma$, and $\beta$. We place uniform priors on each of these parameters. We report the median of the posterior as the parameter estimate, and the symmetric interval surrounding the median that contains 68.3\% of the posterior distribution as the uncertainty. The value for C is -1.87 $\pm$ 0.13, $\gamma$ is -0.13 $\pm$ 0.03, and $\beta$ is -7.67 $\pm$ 1.2. The fit is shown in the left panel of Figure \ref{fig:ha_fit}.

We investigate the possibility that like flare rate as a function of Rossby number, the flare rate follows a saturated relationship with the average quiescent level of $\ha$ emission. We fit a piece-wise function to equation \ref{eq:PW} that models a saturated flare rate below the critical value at -0.71 angstroms and decreases linearly above the critical value, referred to as $H_2$: 

\begin{equation}\label{eq:PW}
 \rm log~R_{31.5}(x_{\ha}) = 
        \begin{cases}
            C  & x_{\ha} \leq X_{\ha_{c}} \\
            \beta~(x_{\ha} - X_{\ha_c}) + C & x_{\ha} > X_{\ha_{c}}\\
        \end{cases}
\end{equation}

\noindent where C is a constant that sets the saturated value of the flare rate and $\beta$ is the linear slope above the critical EW value of -0.71 angstroms. We use \textit{emcee} to determine the best fitting parameters, assuming uniform priors for C and $\beta$. We find C = -1.43 $\pm$ 0.07 and $\beta$ = -8.74 $\pm$ 1.2. The slopes above the critical values of -0.71 angstroms in each of the two models are consistent with each other within the uncertainties. We show the fit in the right panel of Figure \ref{fig:ha_fit}. We find the Bayes factor of H$_1$ to H$_2$ of 365, indicating a strong preference for M$_1$. Indeed, H$_2$ leaves larger residuals below EW = -0.71 (right panel of Figure \ref{fig:ha_fit}).

Flares are known to enhance chromospheric activity indicators such as $\ha$ \citep{Kruse2010,Hilton2010,Bell2012,Kowalski2013}. Although we use the maximum value of $\ha$, hoping to probe the average quiescent $\ha$ emission of the star, it is possible that our observations of the most active stars like G~7-034, which has an maximum EW measurement of -7.38 $\pm$ 0.56 were taken during a flare. Given our small number of spectra per star, this effect could be responsible for the apparent negative slope below the critical value of $\ha$ (Equation \ref{eq:PW_3}). Future work will investigate whether H$_1$ results from flare contamination by gathering more epochs of these stars.

\section{Discussion and Conclusion}\label{sec:DC}
We used a combination of high-resolution spectroscopic observations from TRES and CHIRON with photometric observations from \MEarth-North and \MEarth-South and \tess~to illuminate the relationship between chromospheric activity, stellar rotation, and stellar flare rates for all known single mid-to-late M dwarfs within 15 parsecs and falling within the \tess~Year 1 survey of the southern ecliptic. We measure flare rates or upper limits and equivalent widths of the chromospheric activity tracers He I D$_3$, $\ha$, and the Ca II infrared triplet line at 8542.09 $\angstrom$ for all 125 stars in our sample. We measure 18 new rotation periods spanning 65 $-$ 180 days with \MEarth~and 19 new rotation periods spanning 0.17 $-$ 5.07 days with \tess.  We find that mid-to-late M dwarfs in the solar neighborhood fall into two groups: 26\% have $\ha$ in emission, a saturated flare rate of log R$_{31.5}$ =  -1.30 $\pm$ 0.08 flares per day, and have Rossby numbers less than 0.50; all 32 stars in this regime have a measured rotation period. The remaining 74\% show little to no $\ha$ in emission and exhibit a log flare rate less than -3.86 $\pm$ 0.77 flares per day with the majority of these stars not showing a single flare during the \tess~observations. Out of the 93 stars in the second group, 40 have a measured rotation period; we expect the remaining 53 stars to have rotation periods exceeding 100 days. We find that 60\% of all single mid-to-late M dwarfs in our sample flare one or more times for a total of 1392 flare events.

For each star that flared 5 or more times above the 30\% completeness threshold during the \tess~observations, we fit the flare frequency distribution to determine the slope, $\alpha$.  We find that all stars  exhibit a mutually consistent value of $\alpha$, and we find the statistically weighted mean value to be 1.98 $\pm$ 0.02.  Literature values of $\alpha$ for fully convective M dwarfs observed with space based missions like \kepler~and \ktwo~are roughly consistent with the value of 2 which indicates that the underlying physical mechanism from which the distribution of flares arises may be the same and that micro-flaring could be the mechanism responsible for coronal heating as discussed in \citet{Ilin2019}. \citet{Hawley2014} finds $\alpha$ = 2.01, 2.32 for a sample of 2 mid-to-late M dwarfs. \citet{Silverberg2016} finds $\alpha$ = 2.01 in a comprehensive study of all one-minute cadence \kepler data for one mid M dwarf, GJ~1243. \citet{Davenport2016} finds $\alpha$ = 1.98 for the same star using all \kepler~long- and short-cadence data. \citet{Ilin2019} finds that $\alpha$ in the range of 2.0$-$2.4 for a sample of late-K to mid-M dwarfs. Our value is consistent with, but more precisely determined than these earlier studies. However, this value of 2 is not consistent with flare studies using ground-based data sets like those presented in \citet[][and references there in]{Ramsay2013}. Ground based data sets largely measure values of $\alpha$ less than 2 leading to shallower observed FFDs. As \citet{Hawley2014} points out, this is likely due to better flare statistics obtained with uninterrupted continuous space-based observations that cannot be attained with ground-based observations.

We use this standard value of $\alpha$ to determine a flare rate, R$_{31.5}$, which is the rate of flares above an energy of 3.16$\times$10$^{31}$ ergs. We explore the relationship between flare rate and the EW of $\ha$. We find an increase in the flare rate below a critical value of the $\ha$ equivalent width and a sharp decrease above this critical value (H$_1$). We determine the critical value X$_{\ha_c}$ to be -0.71 angstroms.  We find the slope, $\gamma$ = -0.13 $\pm$ 0.03, below X$_{\ha_c}$, and the slope $\beta$ = -7.67 $\pm$ 1.2 above X$_{\ha_c}$.  We compare these results to a model, H$_2$, showing a saturated relationship below the critical EW value to explore whether flare rate as a function of $\ha$ EW follows the same type of saturated/unsaturated behavior that is observed in flare rate as a function of Rossby number.  We find H$_1$ is strongly favored over H$_2$. Thus Equation \ref{eq:PW_3} allows for the estimation of a flare rate using a measurement of the quiescent $\ha$ EW. This is useful given that flare rates can only be determined accurately by short-cadence continuous observations like those obtained during the \tess~mission, but by comparison, spectra are much easier to obtain and can now provide an estimate for the flare rate of a given mid-to-late M dwarf. Although we used high-resolution spectra, $\ha$ equivalent widths can be readily measured from low-resolution spectrographs as well \citep[e.g.][]{Newton2017}.

There exists an intimate relationship between flares and $\ha$. Not only are flares observed to temporarily enhance the $\ha$ feature, but stars that flare tend to show quiescent $\ha$ in emission. \citet{Kowalski2009}, in a photometric and spectroscopic study of M-dwarfs, finds that stars exhibiting flaring behavior also show $\ha$ in emission in quiescence, consistent with the results of this study. Quiescent $\ha$ is produced via magnetic heating of the chromosphere. During a flare, as the stressed magnetic field loop reconfigures to a lower energy state, energy is transferred leading to observed radiation across the electromagnetic spectrum. Some of that energy goes into enhancing the $\ha$ feature \citep{Kruse2010,Hilton2010,Bell2012,Kowalski2013}. In this work, we find that the flare rate increases as the emission of $\ha$ increases. This result implies that more frequent flaring, even those at low energies, leads to an overall increase in the quiescent $\ha$ equivalent width. However, it could also be that in some cases, each spectrum of a star was contaminated by a large flare, this would artificially enhance the average amount of stellar $\ha$ emission. This could result in the apparent continued relationship between flare rate and $\ha$ below the critical value. We plan to explore this further by obtaining more spectra of the most active stars to better constrain their quiescent amount of $\ha$ emission.

We examine the relationships between the flare rate, and rotation period or Rossby number.  We show that the flare rate displays a saturated relationship below a critical value of the Rossby number much like what has been observed studies of coronal x-ray emission, and chromospheric Ca II H+K, and $\ha$ emission. However, \citet{Lurie2015} shows that the flare rate is not consistent with one single value for three rapidly rotating stars, GJ~1243, GJ~1245A, GJ~1245B when examining the total luminosity emitted in flares normalized by the luminosity of the star in the \kepler~bandpass, $L{fl}/L_{Kp}$. They find that Log $L{fl}/L_{Kp}$ = -3.78, -3.93, and -4.00 for GJ~1243, GJ~1245A, and GJ~1245B respectively. We attribute this difference in flare rate to the stars having different levels of quiescent $\ha$ emission, -4.94 $\angstrom$ \citep{Gizis2002}, -3.70 $\angstrom$ \citep{Mohanty2003}, and -2.5 $\angstrom$ \citep{Browning2010} for GJ~1243, GJ~1245A and GJ~1245B respectively. The rates found for these three stars in \citet{Lurie2015} are consistent with the results we find in this study shown in Figure \ref{fig:ha_fit} where the flare rate increases modestly with increasing $\ha$ emission. \citet{Davenport2016} also examines the relationship between flaring behavior and Rossby number for Kepler (G8-M4)V flare stars. They find that a single power-law model is favored over the piece-wise saturated and unsaturated behavior shown in Figure \ref{fig:prot_fl}. Their sample largely comprises stars that, given their spectral type, would have already transitioned to the unsaturated regime. 

Understanding the flaring behavior and how it evolves throughout the lifetime of fully convective M dwarfs is increasingly important as we try to develop models for how they generate and sustain their magnetic dynamos without the use of a tachocline. As the star ages and spins down, it is postulated that the efficiency of the dynamo decreases leading to a decrease in magnetic activity. We confirm this idea in terms of observed flaring behavior. Several studies posit that magnetic activity can be used as proxy for age because of this intimate relationship between age, rotation, and activity \citep[and references therein]{Kowalski2009,Ilin2019,Davenport2019}. However this relationship may be more complicated than a monotonic decrease over time due to the complexities surrounding how stars shed angular momentum. Observations of open clusters and field stars reveal a bi-modal distribution of rotation periods at a given color used as a proxy for stellar mass. \citet{Redbull2016} finds the presence of two rotational sequences for FGK and early-mid M dwarfs in the 125 Myr open cluster, the Pleiades; a fast- and slow-rotator sequence. \cite{Meibom2011b} showed for the 220 Myr old open cluster M34, that the rotation periods measured for FGK stars also follow two-sequences. \citet{Redbull2017} and \citet{Douglas2016} find a similar bi-modal structure in period-color diagrams for FGK and mid M stars up to the ages of Praespe and Hyades, which are believed to be coeval clusters with an age of $\approx$ 600 Myr. \citet{Newton2016,Newton2018} find that, for M dwarfs, the two rotational sequences are also observed in field stars. \citet{Brown2014} argues that the cause for the bi-modality is two regimes of angular momentum loss.  At early times, the stellar dynamo is weakly coupled to the stellar wind making angular momentum losses small. Angular momentum losses increase as the dynamo transitions abruptly to become strongly coupled to the stellar wind at a later time. \citet{Garraffo2016,Garraffo2018} proposes that the efficiency of angular momentum loss is regulated by magnetic field complexity. They argue that the rearrangement from a multipolar to a dipolar field is the physical mechanism responsible for strong coupling and rapid loss of angular momentum, leading the stars to spin down.  These ideas from \citet{Brown2014} and \citet{Garraffo2016, Garraffo2018} imply that although magnetic activity can be used as a proxy for relative stellar age, i.e. stars with higher levels of magnetic activity are younger than stars with low levels of magnetic activity, it will be difficult to assign more precise stellar ages if there is a stochastic component to the age at which such stars abruptly reconfigure their magnetic fields and begin rapid spin down.

The two different regimes of angular momentum loss may also explain the saturated flaring behavior we observe at Rossby numbers less than 0.5. \citet{Wright2011} notes that the observed behavior in saturated and unsaturated coronal x-ray emission is likely a probe of two different dynamo configurations; one leading to closed magnetic field lines and perhaps a higher magnetic complexity \citep[e.g.][]{Garraffo2016,Garraffo2018} and the other to an increase in open magnetic field lines. Open field lines lead to angular momentum losses via magnetized stellar winds and spin down. In this picture, the saturated stars have a dynamo that is weakly coupled to the stellar wind, produces a multipolar field, and has many closed loops, providing many opportunities for magnetic reconnection. The unsaturated stars have dipolar fields, with more open field lines, and a dynamo that is strongly coupled to the stellar wind leading to angular momentum losses and spin down. Our study provides another link in the study of stellar dynamos and their relation to different geometries of the stellar magnetic field.

The only terrestrial exoplanets for which we will be able to undertake studies of their atmospheres in the next decade will be those found to orbit nearby mid-to-late M-dwarfs. A key input to interpret those observations will be the present levels of flares of the host star, as well as a reasonable understanding of the past history of such flares throughout the planetary lifetime. Flaring activity may alter and erode planetary atmospheres \citep[e.g.][]{Lammer2007,Segura2010,Venot2016,Lingam2017,Tilly2019}, but it may also deliver the UV radiation necessary to facilitate prebiotic chemistry processes \citep{Ranjan2017,Rimmer2018}; UV radiation that is an otherwise negligible component of total bolometric output of these stars. Our study provides both a means to estimate flare rates based on $\ha$ emission, as well as a direct measurement of the flare rates for four of the closest low-mass stars known to host transiting terrestrial planets.

In future work we plan to double the number of targets by incorporating nearby mid-to-late M dwarfs in the northern ecliptic observed during year 2 of the primary mission of \tess. In this follow-up paper, we will also include a study of galactic kinematics as a proxy for age.

\acknowledgments{The authors would like to thanks the anonymous referee for their helpful comments. The authors would also like to thank Elisabeth Newton, Sarah Schmidt, Zachory Berta-Thompson, Jason Dittman, Sebastian Pineda, and Nicholas Mondrik for their contributions to \tess~guest investigator proposal G011231. This research has been supported by RECONS ({\it www.recons.org}) members 
Todd Henry, Hodari James, Leonardo Paredes, and Wei-Chun Jao, who provided data as part of the CHIRON program on the CTIO/SMARTS 1.5m, which is operated as part of the SMARTS Consortium. The authors also thank  Jessica Mink, Gilbert Esquerdo, Perry Berlind, and Michael Calkins for collection and reduction of the TRES data. AAM is supported by NSF Graduate Research Fellowship Grant No. DGE1745303. This work is made possible by a grant from the John Templeton Foundation. The opinions expressed in this publication are those of the authors and do not necessarily reflect the views of the John Templeton Foundation. The MEarth Team gratefully acknowledges funding from the David and Lucile Packard Fellowship for Science and Engineering (awarded to D.C.). This material is based upon work supported by the National Science Foundation under grants AST-0807690, AST-1109468, AST-1004488 (Alan T. Waterman Award), and AST-1616624. This material is based upon work supported by the National Aeronautics and Space Administration under grant 80NSSC19K0635 in support of the TESS Cycle 1 Guest Investigator program, and grant 80NSSC18K0476 issued through the XRP program. This paper includes data collected by the TESS mission, which are publicly available from the Mikulski Archive for Space Telescopes (MAST). This work has made use of data from the European Space Agency mission Gaia (https://www.cosmos.esa.int/gaia), processed by the Gaia Data Processing and Analysis Consortium (DPAC, https://www.cosmos.esa.int/web/gaia/dpac/consortium). Funding for the DPAC has been provided by national institutions, in particular the institutions participat-ing in the Gaia Multilateral Agreement. This work has used data products from the Two Micron All Sky Survey, which is a joint project of the University of Massachusetts and the Infrared Processing and Analysis Center at the California Institute of Technology, funded by NASA and NSF.} 

\vspace{5mm}
\facilities{TESS, MEarth, FLWO:1.5m (TRES), CTIO/SMARTS:1.5m (CHIRON)} 

\software{{\sc celerite} \citep{Foreman-Mackey(2017)}, {\sc exoplanet} \citep{exoplanet:exoplanet}, {\sc PYMC3} \citep{exoplanet:pymc3}, {\sc python}}

This research made use of {\sc exoplanet} \citep{exoplanet:exoplanet} and its dependencies \citep{exoplanet:astropy13, exoplanet:astropy18, exoplanet:exoplanet, exoplanet:foremanmackey17, exoplanet:foremanmackey18, exoplanet:kipping13, exoplanet:luger18, exoplanet:pymc3, exoplanet:theano}.

\bibliographystyle{aasjournal}
\bibliography{references}{}

\end{document}